\documentclass[12pt]{article}
\usepackage{amsmath}
\usepackage{amssymb}
\begin{document}
\begin{center}
{\large\bf  {FLPR Model: Nilpotent (Anti-)co-BRST Symmetries }}

\vskip 3.0cm

{\sf R. P. Malik$^{(a,b)}$}\\
$^{(a)}$ {\it Physics Department, Institute of Science,}\\
{\it Banaras Hindu University, Varanasi - 221 005, (U.P.), India}\\

\vskip 0.1cm

\vskip 0.1cm

$^{(b)}$ {\it DST Centre for Interdisciplinary Mathematical Sciences,}\\
{\it Institute of Science, Banaras Hindu University, Varanasi - 221 005, India}\\
{\small {\sf {e-mails: rpmalik1995@gmail.com; malik@bhu.ac.in}}}

\end{center}

\vskip 1cm

\noindent
{\bf Abstract:}
We demonstrate the existence of a set of {\it novel} off-shell nilpotent and absolutely anticommuting
continuous symmetry transformations, within the framework of the Becchi-Rouet-Stora-Tyutin (BRST)
formalism, which are over and above the {\it usual} off-shell nilpotent and absolutely anticommuting
(anti-)BRST transformations that are respected by the {\it quantum} version of the {\it classical} first-order Lagrangian for the
Friedberg-Lee-Pang-Ren (FLPR)  model that describes the motion of a single non-relativistic particle
of unit mass (moving under the influence of a general rotationally invariant {\it spatial} two-dimensional  
potential). We christen these novel set of  fermionic transformations as the (anti-)co-BRST
transformations because the gauge-fixing terms remain invariant under them. 
We derive the conserved and off-shell nilpotent (anti-)BRST and (anti-)co-BRST charges and 
comment on the physicality criteria w.r.t. them where we establish the presence of the operator
forms of the first-class constraints (of the original {\it classical} theory) at the {\it quantum} level.





\newpage

\section{Introduction}

The Friedberg-Lee-Pang-Ren (FLPR) model [1] is a U(1) solvable gauge system of a single non-relativistic particle 
of unit mass that moves under
the influence of the general {\it spatial} 2D rotationally invariant potential where the {\it correct} results have been obtained by taking into
account {\it all} the Gribov-type gauge-equivalent copies which are different from the {\it one} that was suggested by Gribov
himself in his seminal paper [2]. Basically, the solvable gauge invariant FLPR model is inspired by the dimensional 
reduction of the ordinary Yang-Mills theories to the one (0 + 1)-dimensional (1D) quantum mechanical system (see, e.g. [1] for details).
In our present endeavor, we apply the theoretical beauty and strength of the 
Becchi-Rouet-Stora-Tyutin (BRST) formalism to {\it this} model and BRST-quantize the corresponding theory by generalizing its
{\it classical} infinitesimal and local gauge transformations to their counterparts {\it quantum} 
off-shell nilpotent and absolutely anticommuting (anti-)BRST transformations under which the 
total kinetic terms of the Lagrangian (for the FLPR model) remain invariant. Over and above the 
gauge and nilpotent (anti-)BRST transformations, we demonstrate the existence of the 
off-shell nilpotent and absolutely anticommuting (anti-)dual BRST [i.e. (anti-)co-BRST] symmetry transformations
under which the total gauge-fixing terms of our present theory remain invariant. We mention, in passing, a {\it unique}
bosonic symmetry transformation under which the Faddeev-Popov (FP) ghost terms of our BRST-quantized FLPR theory remain invariant. 
This unique bosonic symmetry (i) commutes with all the {\it four} nilpotent (anti-)BRST and (anti-)co-BRST
transformations, and (ii) is a {\it unique} anticommutator between the fermionic (anti-)BRST
and (anti-)co-BRST transformation operators. Future scope, significance and perspective of {\it all} these fermionic and bosonic symmetry
transformations are pointed out in a concise manner.

Our present endeavor is essential and important on the following counts. First, if the beauty of a theory is defined in 
terms of the number of continuous and discrete symmetry transformations it respects, the models of Hodge theory 
(see, e.g. [3,4] and references therein) belong to this category. This is due 
to the fact that such models, within the framework of BRST formalism, respect {\it six} continuous symmetries
and, at least,  a couple of discrete duality symmetry transformations where the most {\it basic} continuous transformations are the
nilpotent (i.e. fermionic) (anti-)BRST and (anti-) co-BRST transformations.  In our present endeavor, we have concentrated on the 
existence of the (anti-)co-BRST transformations over and above the {\it usual} (anti-)BRST transformations for the
FLPR model. The {\it latter} symmetry transformations have
 been pointed out in an earlier work [5].  Second, the FLPR model has been studied by many authors from different theoretical angles 
(see. e.g. [6,7] and references therein). It is, therefore, very interesting to say something {\it new} about it and prove it to be a tractable
1D toy model for the Hodge theory (see, e.g. [8]). Third,
the ideas of the symmetry transformations, related conserved charges and their associated algebra, etc., have played a pivotal role
in the modern developments of theoretical (high energy) physics. Hence, it is always interesting to find out the {\it novel}
symmetries for a given theoretically useful model of physical interest. In our present
investigation, we focus on a set of novel fermionic symmetries 
at the {\it quantum} level that are over and above (i) the {\it classical} gauge transformations, and
(ii) the nilpotent {\it quantum} (anti-)BRST transformations [5]. Finally, there have been critical remarks on the {\it independent} nature of the
(anti-)co-BRST transformations {\it but} the mathematical basis for the existence of these symmetries has been 
elaborated (see, e.g. [9] and references therein) which has been highlighted in a nice recent work [10] where a unified framework 
has been provided for {\it all} kinds of BRST-type symmetries and a few {\it new} BRST-type symmetries have been proposed.
We wish to devote time on (i) the derivations of the above {\it new} BRST-type off-shell nilpotent transformations
by exploiting the theoretical strength of the supervariable/superfield approach so that the geometrical basis
for their existence can be provided, and (ii) the off-shell nilpotency and absolute anticommutativity properties 
can be expressed in terms of a few geometrical quantities on the supermanifold.

The theoretical contents of our present endeavor are organized as follows. In section two, 
we provide a brief synopsis of the first-class constraints, ensuing {\it classical}  gauge symmetry transformations and their
elevation to their counterparts {\it quantum} (anti-)BRST symmetry transformations [5]. Our section three deals with the discussion
on the off-shell nilpotent and absolutely anticommuting (anti-)co-BRST symmetry transformations.  Section four of
our present Letter is devoted to a concise discussion on the physicality criteria w.r.t. the nilpotent (anti-)BRST
and (anti-)co-BRST charges. Finally, in section five, we make some concluding remarks and point out the future perspective
of our present endeavor.


\section{Preliminaries: Gauge and (anti-)BRST symmetries}

\noindent
We begin with the first-order Lagrangian ($L_f$) for the FLPR model that describes the motion
of a single non-relativistic particle (with unit mass) that is moving under the influence of a general {\it spatial} 2D rotationally 
invariant potential of the form: $U (x^2 + y^2)$. This Lagrangian, in the Cartesian coordinate system,  
is as follows\footnote{Our first-order Lagrangian
($L_f$) is {\it equal} to the second-order FLPR Lagrangian ($L_0$) if we make use of the EL-EoMs: 
$\dot x - p_x + g\,\zeta\, y = 0, \, \dot y - p_y - g\,\zeta\, x = 0,\, \dot z - p_z - \zeta = 0$  that are derived from $L_f$ to re-express $L_f$ as: 
$L_0 = \frac{1}{2}\, \bigl [ (\dot x + g\, \zeta\, y)^2 + (\dot y - g\, \zeta\, x)^2 + (\dot z - \zeta)^2 \bigr ] - U (x^2 + y^2)$
which is the {\it original} Lagrangian that was proposed by FLPR [1]. In our present endeavor, we shall concentrate on the
first-order Lagrangian ($L_f$) because it incorporates into itself more number of variables than the second-order Lagrangian ($L_0$). Hence,
theoretically, the first-order Lagrangian ($L_f$) is more  interesting and appealing [11]. It is worthwhile to point out that
in [1], the 1D trajectory of the non-relativistic particle is assumed to be embedded in the 
target space of the {\it four} dimensional configuration
space where ($x, y, z, \zeta$) are treated on equal footing as the Cartesian coordinates. The {\it latter}  are, of course,  function of 
the evolution parameter $t$.}
(see, e.g. [5] for details)
\begin{eqnarray}
L_{f} = p_x \, \dot x + p_y \, \dot y + p_z \, \dot z - \frac{1}{2}\, \big (p_x^2 + p_y^2 + p_z^2 \big ) 
-\zeta\, \bigl [g (x\,p_y - y\, p_x) + p_z \bigr ] - U(x^2 + y^2),
\end{eqnarray}
where $(\dot x, \dot y, \dot z)$ are the generalized velocities 
[with $\dot x = (dx/dt), \dot y = (dy/dt), \dot z = (dz/dt)$]
corresponding to the Cartesian coordinates $(x, y, z)$
of the instantaneous position of the non-relativistic particle whose trajectory is parameterized by the ``time'' evolution 
parameter $t$. The canonical conjugate momenta $(p_x, p_y, p_z)$, corresponding to the coordinates $(x, y, z)$, are
constrained (through the Lagrange multiplier variable
$\zeta (t)$) to satisfy the relationship: $ g \,(x\,p_y - y\, p_x) + p_z \approx 0$ where $g> 0$ is the real positive coupling constant.
As is obvious, all the variables of the Lagrangian (1) are function of the evolution parameter $t$.

It is straightforward to note that the canonical conjugate momentum $ p_\zeta \approx 0$ [with $p_\zeta = (\partial L_f / \partial \dot \zeta)$]
is the primary constraint on the theory in the terminology of Dirac's prescription for the classification 
scheme of constraints (see, e.g. [12,13]). For the derivation of the secondary, tertiary and higher order constraints, the Hamiltonian approach is
the most appropriate one. However, for our simple system (see, e.g. [14]), the following Euler-Lagrange (EL) equation of motion (EoM) derived
from the first-order Lagrangian ($L_f$)
w.r.t. the Lagrange multiplier variable $\zeta (t)$, namely; 
 \begin{eqnarray}
\frac{d}{dt}\, \Bigl (\frac{\partial L_{f}} {\partial \dot \zeta} \Bigr ) = \frac{\partial L_{f}} {\partial \zeta} \quad
\Longrightarrow \quad \dot p_\zeta = -\,\bigl [g\, (x\,p_y - y\, p_x) + p_z \bigr ] \approx 0,
\end{eqnarray} 
implies the time-evolution invariance [14]
of the primary constraint ($p_\zeta \approx 0$) which leads to the derivation of the secondary constraint:
$ -\,\bigl [g \,(x\,p_y - y\, p_x) + p_z \bigr ] \approx 0$. There are {\it no} further constraints on our theory.
 At this stage itself, it can be seen that {\it both} the above
constraints are first-class in the terminology of Dirac's prescription for the classification scheme of constraints  
because they commute with each-other. The {\it latter} statement is true as there is {\it no} presence of 
the variable $\zeta (t)$ in the secondary constraint
and other coordinates (e.g. $x,\, y$) and momenta (e.g. $p_x, \, p_y, \, p_z$) in {\it it} commute with $p_\zeta \approx 0$.

The existence of the first-class constraints (e.g. $p_\zeta \approx 0, \, g\, (x\,p_y - y\, p_x) + p_z \approx 0$) on our theory is a sure
signature of our theory being a gauge theory whose 
following local, continuous and infinitesimal {\it classical}\footnote{We differ by an overall sign factor
from the local, continuous and infinitesimal gauge symmetry transformations that have been mentioned in [5] where the
quantization of the FLPR model has been performed by using the 
{\it modified} Faddeev-Jackiw formalism (MFJF) of the reduced phase space [15].
The connected algorithm for the validity of the MFJF has been proposed in a nice piece of work [16].} gauge symmetry transformations ($\delta_g$) 
 \begin{eqnarray}
 \delta_g x &=& -\, g\, y \,\lambda, \quad \delta_g y = g \,x \,\lambda, \quad \delta_g z = \lambda, \quad \delta_g p_x = -\, g\, p_y\, \lambda, \nonumber\\
\delta_g p_y &=& g\,p_x \,\lambda, \quad \delta_g p_z = 0, 
\quad \delta_g p_\zeta  = 0, \quad \delta_g \zeta  = \dot \lambda, \quad (\delta_g L_f = 0),    
\end{eqnarray}
are generated by the generator ($G$) that can be precisely
expressed in terms the above first-class constraints of our theory as (see, e.g. [17] for details)
\begin{eqnarray}
G = \dot \lambda\, p_\zeta + \lambda\, [g\, (x\, p_y - y\, p_x) + p_z],
\end{eqnarray}
where $ \lambda (t)$ is the infinitesimal gauge symmetry transformation parameter. It is straightforward to note that the above generator leads to
the derivation of the classical gauge symmetry transformations ($\delta_g$) if we use the standard relationship between the infinitesimal continuous
gauge symmetry transformation ($\delta_g$)
for the generic variable $\phi (t)$ of our theory [cf. Eq. (1)]
and the generator $G$ [cf. Eq. (4)] as 
\begin{eqnarray}
\delta_g \, \phi (t) = -i\, [ \phi (t), \;G], \qquad \phi = x, y, z, \zeta, p_x, p_y, p_z, p_\zeta,
\end{eqnarray}
provided we exploit the standard non-zero canonical commutators\footnote{For the sake of brevity and the relevance of the
{\it classical} gauge symmetry transformations (3) in the context of the forthcoming {\it quantum} (anti-)BRST symmetry transformations,
we have used the canonical commutators for the derivation of the infinitesimal gauge symmetry transformations (3). However, at the
{\it classical} level, we can very well use the canonical Poisson brackets (instead of the canonical commutators) which would lead to the derivation
of the infinitesimal gauge symmetry transformations (3) provided we take care of the presence of the appropriate $i$ factors. In other words,
we can always replace: $-i\, [ \phi (t), \;G]$ of the r.h.s. of (5) by the Poisson bracket: $ \{ \phi (t), \;G\}_{PB}$ where the notation
$ \{.., \;..\}_{PB}$ stands for the definition of the canonical Poisson bracket in the phase space of the system under consideration.}
(in the natural units where $\hbar = c = 1$). For instance, we have the non-zero commutators as:  $ [x, \,p_x] = i, \,\,[y,\, p_y] = i,\,\,
[z, \, p_z] = i,\,\,[\zeta, \, p_\zeta] = i$ for our theory. All the rest of the canonical commutators are 
zero by definition according to the rules that are set by the canonical quantization scheme.

The above classical infinitesimal gauge symmetry transformations ($\delta_g$) can be elevated to their quantum counterparts 
infinitesimal, continuous and off-shell nilpotent ($s_{(a)b}^2 = 0$) (anti-)BRST symmetry transformations
[$s_{(a)b}$] as 
\begin{eqnarray}
 s_{ab} \, x &=& -\, g\, y \,\bar c, \quad s_{ab}\, y = g \,x \,\bar c, \quad s_{ab}\, z = \bar c, \quad s_{ab}\, \zeta  = \dot {\bar c},
\quad s_{ab}\, p_x = -\, g\, p_y\, \bar c, \nonumber\\
s_{ab}\, p_y &=&  g \,p_x \,\bar c, \quad s_{ab}\, p_z = 0,
\quad s_{ab}\, p_\zeta  = 0,  \quad s_{ab} \,\bar c = 0, 
\quad s_{ab}\,c = - i\, b, \quad s_{ab}\,b = 0, \nonumber\\ 
 s_{b} \, x &=& -\, g\, y \,c, \quad s_{b}\, y = g \,x \, c, \quad s_{b}\, z =  c, \quad s_{b}\, \zeta  = \dot  c,
\quad s_{b}\, p_x = -\, g\, p_y\,  c, \nonumber\\
s_{b}\, p_y &=&  g \,p_x \, c, \quad s_{b}\, p_z = 0,
\quad s_{b}\, p_\zeta  = 0,  \quad s_{b} \, c = 0, 
\quad s_{b}\,\bar c =  i\, b, \quad s_{b}\,b = 0, 
\end{eqnarray}
which are the {\it symmetry} transformations for the generalized version of the first-order classical version of the Lagrangian ($L_f$)
to its counterpart quantum (anti-)BRST invariant Lagrangian ($L_b$) that incorporates into itself the 't Hooft like [18]
 gauge-fixing term  (used in the context of (i) the Abelian Higgs model, and/or (ii) the St$\ddot u$ckelberg-modified massive Abelian 
1-form gauge theory) {\it and} the fermionic Faddeev-Popov (FP) ghost terms as: 
\begin{eqnarray}
L_{b} &=& p_x \, \dot x + p_y \, \dot y + p_z \, \dot z - \frac{1}{2}\, \big (p_x^2 + p_y^2 + p_z^2 \big ) 
-\zeta\, \bigl [g (x\,p_y - y\, p_x) + p_z \bigr ] - U(x^2 + y^2) \nonumber\\
&+& b \,\bigl (\dot \zeta - z \bigr ) + \frac{1}{2}\, b^2 - i \, \dot {\bar c}\, \dot c -\, i\,\bar c\, c. 
\end{eqnarray}
We observe the following transformations of (7) under the above (anti-)BRST symmetries
\begin{eqnarray}
s_{ab} \, L_b = \frac{d}{dt} \, \Big (b\, \dot {\bar c} \Big ), \qquad  s_{b} \, L_b = \frac{d}{dt} \, \Big (b\,  \dot c \Big ),
\end{eqnarray}
which establish the (anti-)BRST invariance of the action integral corresponding to the Lagrangian $L_b$. In the equations (6) and (7),
the (anti-)ghost variables $(\bar c)c$ are fermionic ($c^2 = \bar c^2 = 0, c\, \bar c + \bar c \, c= 0$) in nature which are
invoked  to maintain the unitarity in the theory and $b$ is the Nakanishi-Lautrup type auxiliary variable that is needed
to linearize the {\it original} quadratic gauge-fixing term [i.e. $  b\, (\dot \zeta - z) + (b^2/2) \equiv - (1/2)\, (\dot \zeta - z)^2 $].

We end this section with the following crucial comments. First, the nilpotency property shows that the (anti-)BRST symmetry  transformations
are fermionic in nature and, therefore, these symmetries transform a bosonic variable into fermionic variable and vice-versa. Second, these
symmetry transformations turn out to be absolutely anticommuting (i.e. $s_b \,s_{ab} + s_{ab}\, s_b = 0$) in nature which establishes the 
linear independence of the BRST and anti-BRST symmetry transformations. Third, the first order Lagrangian $L_f$ [cf. Eq. (1)], being gauge invariant, is
also (anti-)BRST invariant (i.e. $s_{(a)b} \, L_f = 0$) and the total kinetic terms of our theory remain invariant under the (anti-)BRST
symmetry transformations. Four, the gauge  as well as the (anti-)BRST symmetry transformations 
[cf. Eqs. (3),(6)] on the Lagrange multiplier variable establish
it as the ``gauge'' variable in our theory.  
Fifth, we note that the gauge-fixing and FP-ghost terms of our (anti-)BRST invariant theory, described by
the Lagrangian $L_b$ [cf. Eq. (7)], can be expressed (modulo some total time derivatives) in {\it three} different ways because of the following observations
\begin{eqnarray}
L_b &=& L_f + s_b \bigl [-\, i\, \bar c \,\{(\dot \zeta - z) + \frac{1}{2}\, b \} \bigr ]
    \equiv L_f + s_{ab} \bigl [+\, i\, c \, \{(\dot \zeta - z) + \frac{1}{2}\, b \} \bigr ], \nonumber\\
    &\equiv& L_f + s_b s_{ab} \bigl [\, \frac{i}{2}\, \zeta^2  + \frac{i}{2}\, z^2 - \frac{1}{2}\, \bar c\, c \bigr ],
\end{eqnarray}
which establish the (anti-)BRST invariance (modulo some total time derivatives)
of the Lagrangian $L_b$ due to the off-shell nilpotency (i.e. $s_{(a)b}^2 = 0$) of the (anti-)BRST  symmetry transformations
and the (anti-)BRST invariance of the first-order Lagrangian. Finally, the 
infinitesimal (anti-)BRST symmetry transformations (6) are generated by the following
conserved (anti-)BRST charges (derived from the celebrated Noether theorem), namely;
\begin{eqnarray}
Q_{ab} &=& \bigl [g \, (x\, p_y - y \, p_x) + p_z \bigr ]\, \bar c + b \, \dot {\bar c} 
 \equiv b\, \dot {\bar c} - \dot b \, \bar c, \nonumber\\
Q_{b} &=& \bigl [g \, (x\, p_y - y \, p_x) + p_z \bigr ]\,  c + b \, \dot {c} 
 \equiv b\, \dot {c} - \dot b \, c, 
\end{eqnarray}
where we have used the EL-EoM: $\,\dot b = -\, \bigl [g \, (x\, p_y - y \, p_x) + p_z \bigr ]$ w.r.t. the gauge variable $\zeta (t)$
to obtain the concise forms of the (anti-)BRST charges. The conservation laws (i.e. $\dot Q_{(a)b} = 0 $) for the (anti-)BRST charges 
can be proven by using the EL-EoMs: $\ddot c = c, \;  \ddot {\bar c} = {\bar c}, \; \ddot b = b$. The derivation of the last entry
 (i.e. $\ddot b = b$)
requires  many other EL-EoMs [see, e.g. Eq. (14) below] from the Lagrangian $L_b$ which we discuss in our next section.


\section{(Anti-)co-BRST symmetry transformations}

\noindent
The (anti-)BRST invariant Lagrangian $L_b$ [cf. Eq. (7)], in addition to the (anti-)BRST symmetry transformations, respects another
set of off-shell nilpotent (i.e. $s_{(a)d}^2 = 0$) and absolutely anticommuting (i.e. $s_d \, s_{ad} + s_{ad}\, s_d = 0$) (anti-)dual-BRST
symmetry transformations. In literature, these {\it fermionic} symmetries are 
also christened as the (anti-)co-BRST symmetry transformations under
which the gauge-fixing terms remain invariant. For our theory under consideration
(described by the Lagrangian $L_b$), these infinitesimal, continuous
and off-shell nilpotent (anti-)co-BRST symmetry transformations ($s_{(a)d}$) are:
\begin{eqnarray}
 s_{ad} \, x &=& -\, g\, y \,\dot {c}, \quad s_{ad}\, y = g \,x \,\dot c, \quad s_{ad}\, z = \dot c, \quad s_{ad}\, \zeta  = c,
\quad s_{ad}\, p_x = -\, g\, p_y\, \dot c, \quad s_{ad}\,c = 0, \nonumber\\
s_{ad}\, p_y &=&  g \,p_x \,\dot c, \quad s_{ad}\, p_z = 0,
\quad s_{ad}\, p_\zeta  = 0,  \quad s_{ad} \,\bar c = i\, \bigl [g (x\, p_y - y\, p_x) + p_z \bigr ], \quad s_{ad}\,b = 0, 
\nonumber\\ 
 s_{d} \, x &=& -\, g\, y \,\dot {\bar c}, \quad s_{d}\, y = g \,x \,\dot {\bar c} , \quad s_{d}\, z = \dot {\bar  c}, 
\quad s_{d}\, \zeta  = \bar c,
\quad s_{d}\, p_x = -\, g\, p_y\, \dot {\bar c}, \quad s_{d}\,b = 0, \nonumber\\
s_{d}\, p_y &=&  g \,p_x \, \dot {\bar c}, \quad s_{d}\, p_z = 0,
\quad s_{d}\, p_\zeta  = 0,  \quad s_{d} \, \bar c = 0, 
\quad s_{d}\, c = - \, i\, \bigl [g (x\, p_y - y\, p_x) + p_z \bigr ].   
\end{eqnarray}
It can be readily verified, from the above transformations that we have: $s_{(a)d}\, [(\dot \zeta - z)]  = 0, \, s_{(a)d}\, b = 0$ 
which imply that the {\it total} gauge-fixing terms remain invariant.

Under the above infinitesimal, nilpotent and absolutely anticommuting 
 (anti-)co-BRST symmetry transformations, it is an elementary exercise to check that the (anti-)BRST invariant Lagrangian
$L_b$ transforms to the total time derivatives as
\begin{eqnarray}
s_{ad}\, L_b = \frac{d}{dt}\, \Bigl [ \{g \, (x\, p_y - y \, p_x) + p_z \}\, \dot c \Bigr ], \quad
s_{d}\, L_b = \frac{d}{dt}\, \Bigl [ \{g \, (x\, p_y - y \, p_x) + p_z \}\, \dot {\bar c} \Bigr ],
\end{eqnarray}
which demonstrate that the action integral $S = \int dt\, L_b$ remains invariant because {\it all} the physical variables and the (anti-)ghost 
variables vanish off as $ t \to \pm \, \infty$. According to the  Noether theorem, the above observations of the invariance of the action 
integral (corresponding to the (anti-)BRST and (anti-)co-BRST invariant Lagrangian $L_b$)
lead to the derivations of the following (anti-)co-BRST charges ($Q_{(a)d}$), namely;
\begin{eqnarray}
Q_{ad} &=& \bigl [g\, (x\, p_y - y \, p_x) + p_z \bigr ] \, \dot c + b\, c \equiv b\, c - \dot b\, \dot c, \nonumber\\  
Q_{d} &=& \bigl [g\, (x\, p_y - y \, p_x) + p_z \bigr ] \, \dot {\bar c} + b\, \bar c \equiv b\, {\bar c} - \dot b\, \dot {\bar c}.
\end{eqnarray}
The conservation of the above charges can be proven by using the appropriate EL-EoMs that are derived from the Lagrangian $L_b$. For
instance, we can check, in a straightforward manner, that the following EL-EoMs
\begin{eqnarray}
&&\dot x = p_x - g\,\zeta\,y, \qquad \dot y = p_x + g\,\zeta\,x, \qquad \dot z = p_z + \zeta\, \qquad \dot p_z = -\, b, \nonumber\\
&&\dot p_x =- g\,\zeta\,p_y -\, 2\,x\, U^\prime, \quad \dot p_y = g\, \zeta\,p_x  -\, 2\,y\, U^\prime, \quad 
\dot b = -\, \bigl [g\, (x\, p_y - y \, p_x) + p_z \bigr ],
\end{eqnarray}
are useful in the derivation: $\ddot b = b$ which turns out to play a crucial role in the proof that
the above (anti-)co-BRST charges are conserved (i.e. $\dot Q_{(a)d} = 0$) with the additional help 
from the EL-EoMs: $\ddot c = c, \; \ddot {\bar c} = {\bar c}$.  
It is essential to point out the fact that the symbol $U^\prime$ in (14) stands for the derivative of the spatially 
2D rotationally invariant general potential [of the form:  $U(x^2 + y^2)$]  w.r.t. its full argument: $(x^2 + y^2)$.

We conclude this section with the following crucial remarks. First, as pointed out earlier, we note that that the total kinetic terms 
of the Lagrangian $L_b$ 
of our theory (describing FLPR model) remain invariant under the (anti-)BRST symmetry transformations. On the other hand,
it is the total gauge-fixing terms that remain invariant under the (anti-)co-BRST symmetry  transformations. In the usual D-dimensional ($D\geq 2$) 
Abelian $p$-form (with $p = 1, 2, 3...$) gauge theories, the kinetic term for the gauge field owes its origin to the exterior derivative 
of differential geometry [19,20]. On the other hand, the gauge-fixing term for the gauge field 
crucially depends on the operation of the co-exterior derivative
on the Abelian $p$-form gauge field. This is why the symmetry invariances of these terms are associated with the nomenclatures of the (anti-)BRST
and (anti-)co-BRST symmetry transformations, respectively. In our present endeavor, we have borrowed  these nomenclatures which are actually valid for
the BRST-quantized D-dimensional ($D\geq 2$) gauge theories (even though, in our case, the
trajectory of the non-relativistic particle is characterized by a 1D evolution parameter $t$). Second, the off-shell nilpotency as well
as the absolute anticommutativity properties of the
(anti-)co-BRST transformations are {\it true} because we observe that: $s_{(a)d} \, \bigl [g (x\, p_y - y\, p_x) + p_z \bigr ] = 0$.
Third, it is very interesting to point out that the first-class constraints of our theory are invariant under (i) the infinitesimal gauge
symmetry transformations (3), (ii) the infinitesimal nilpotent (anti-)BRST symmetry transformations (6), and
(iii) the infinitesimal nilpotent (anti-)co-BRST symmetry transformations (11), too. Hence these constraints are {\it physical} restrictions on
our theory which should be respected at the classical level as well as at the quantum level. Finally, the 
infinitesimal (anti-)BRST as well as the
(anti-)co-BRST transformations are fermionic 
(i.e. nilpotent) type transformations and, as a consequence, they transform the bosonic variables
into their counterparts fermionic variables and vice-versa. However, they are {\it not} like the $\mathcal {N} = 2$ SUSY transformations
because, whereas the (anti-)BRST as well as the (anti-) co-BRST symmetry transformations are absolutely anticommuting in nature, the
same is {\it not} true\footnote{For the $\mathcal {N} = 2$ SUSY quantum mechanical models, the two SUSY
transformations $s_1$ and $s_2$ are nilpotent ($s_1^2 = s_2^2 = 0$) of order two. However, their anticommutator, acting
on a variable, leads to the time-translation of that specific variable which is generated by the Hamiltonian
operator ($H$) of the $\mathcal {N} = 2$ SUSY quantum mechanical models. The $\mathcal {N} = 2$ SUSY charges $Q(\bar Q)$ and $H$
obey the algebra: $ Q^2 = {\bar Q}^2 = 0,\, \{Q, \; \bar Q\} = H, \, [H, \, Q] = [H, \, \bar Q] = 0$ which is {\it also} reminiscent of the
Hodge algebra.}
for the $\mathcal {N} = 2$ SUSY symmetry transformations despite the fact that the {\it latter} symmetry transformations
are {\it also} (i) nilpotent of order two, and (ii) only {\it two} in numbers.


\section{Physicality criteria: Nilpotent charges}

\noindent
Before we discuss the physicality criteria w.r.t. the conserved and nilpotent (anti-)BRST and (anti-)co-BRST charges, we have 
to discuss a few crucial points 
that are essential as the background theoretical material. In this context, first of all, we note that the conserved 
(anti-)BRST and (anti-)co-BRST charges [cf. Eqs. (10),(13)] are the generators for the (anti-)BRST as well as
the (anti-)co-BRST symmetry transformations [cf. Eqs. (6),(11)] which are respected by the Lagrangian $L_b$ [cf. Eq. (7)] that contains
bosonic as well as the fermionic variables. Hence, the general formula between the continuous symmetry transformations and their generators
as the conserved charges gets modified from (5) to the following more general mathematical relationship
\begin{eqnarray}
s_r\, \phi (t) = -i\, \bigl [\phi (t), \; Q_r \bigr ]_{(\pm)} \quad \mbox{with} \quad r = b, ab, d, ad,
\end{eqnarray}
where the ${(\pm)}$ signs, as the subscripts on the square bracket, stand for the bracket to be an (anti)commutator for the the generic variable $\phi (t)$
of the Lagrangian $L_b$ being the fermionic (i.e. $\phi = \bar c,\, c $) and  bosonic (i.e. $\phi = x, y, z, \zeta, p_x, p_y, p_z, b$)
in nature, respectively. To verify the sanctity of the above relationship in the 
context of the infinitesimal (anti-)BRST [cf. Eq. (6)] and (anti-)co-BRST symmetry 
transformations [cf. Eq. (11)], we have to use the non-zero canonical commutators that have been listed after
equation (5) and, in addition, we have to make use of the anticommutators: $\{c (t), \; \dot {\bar c} (t) \} = +\,1, \;
 \{\bar c (t), \; \dot {c} (t) \} = -\,1$. It goes without saying that all the {\it rest} of the (anti)commutators for our theory are
 zero by definition (according the rules that are set by the canonical quantization scheme). It is straightforward to check the usefulness
 and importance of the the above {\it basic} canonical anticommutators in the derivations of
the infinitesimal (anti-)BRST and (anti-)co-BRST symmetry transformations: $s_b \, \bar c = -\, i\, \{\bar c, \; Q_b\} = i\, b, \;
 s_{ab} \, c = -\, i\, \{c, \; Q_{ab}\} = -\, i\, b,\; s_{d} \, c = -\, i \, \{c, \; Q_{d}\} = -\, i\,\bigl [g \,(x\, p_y - y\, p_x) + p_z \bigr ], \;
 s_{ad}\, \bar c = -\,i\, \{\bar c,\; Q_{ad} \} = +\, i\, \bigl [g\, (x\, p_y - y\, p_x) + p_z \bigr ]$. Similarly, the non-zero commutators [that
have been listed after equation (5)] play their decisive roles in the derivations of the nilpotent symmetry transformations for the bosonic variables
of our theory that is described by the {\it quantum} Lagrangian $L_b$ [cf. Eq. (7)].

As far as the physicality criteria w.r.t. the conserved (anti-)BRST and (anti-)co-BRST charges are concerned, it is very urgent 
to prove their nilpotency (i.e. $Q_{(a)b}^2 = Q_{(a)d}^2 = 0$) property because, keeping in our mind the BRST cohomology (see. e.g. [14]), it is
very essential that the above conserved charges {\it must} be nilpotent of order two.
 Towards this goal in mind, we would like to point out that the relationship (15) is very general and it can be exploited to prove
 the nilpotency property of the above conserved charges. In this context, it is straightforward to note that the following relationships
 are {\it true}, namely;
\begin{eqnarray}
s_b\, Q_b &=& -\, i\, \{Q_b, \; Q_b \} = 0, \qquad \qquad s_{ab}\, Q_{ab} = -\, i\, \{Q_{ab}, \; Q_{ab} \} = 0, \nonumber\\
s_d\, Q_d &=& -\, i\,\{Q_d, \; Q_d \} = 0, \qquad \qquad s_{ad}\, Q_{ad} = -\, i\, \{Q_{ad}, \; Q_{ad} \} = 0.
\end{eqnarray}
In the above, the following inputs have been taken into account. First of all, a close look at the expressions for the 
conserved (anti-)BRST and (anti-)co-BRST charges [cf. Eqs. (10),(13)] demonstrate that they are fermionic in nature because
{\it each} term of these expressions has a bosonic part and a fermionic (anti-)ghost variable. Second, we have used the general relationship (15)
between the continuous symmetry transformations and their generators as the conserved charges with special emphasis on the 
nature of the generic variable $\phi (t)$ of our theory that is described by the Lagrangian $L_b$. Finally, we have explicitly computed
the l.h.s. of the above equation (16) by directly applying the transformations (6) and (11) on the conserved charges
that have been listed in (10) and (13), respectively.

It is worthwhile to mention, in passing, that the absolute anticommutativity properties of the above conserved 
and nilpotent charges can {\it also} be proven (in exactly the same manner). For instance, the l.h.s. of 
$s_b\, Q_{ab} = -\, i\,\{ Q_{ab}, \; Q_b \}$ and  $s_d\, Q_{ad} = -\, i\,\{ Q_{ad}, \; Q_d \}$ can be
computed by the direct applications of the nilpotent (anti-)BRST and (anti-)co-BRST symmetry
transformations [cf. Eqs. (6),(11)] on the explicit forms of the (anti-) BRST and (anti-)co-BRST charges
[cf. Eqs. (10),(13)] and can be shown to be equal to {\it zero} which would establish the absolute anticommutativity
(i.e. $Q_b \, Q_{ab} + Q_{ab}\, Q_b = 0$ and  $Q_d \, Q_{ad} + Q_{ad}\, Q_d = 0$) properties of the
(anti-)BRST and (anti-)co-BRST charges, respectively. Similar results can be obtained from the computations of
the l.h.s. of the expressions:
$s_{ab}\, Q_{b} = -\, i\,\{ Q_{b}, \; Q_{ab} \}$ and  $s_{ad}\, Q_{d} = -\, i\,\{ Q_{d}, \; Q_{ad} \}$
by using the symmetry transformations in (6) and (11) and the conserved charges in (10) and (13) which would
amount to the proof of the absolute anticommutativity properties of the above conserved and nilpotent charges.
These observations would physically establish the linear independence of (i) the BRST and anti-BRST
symmetries and the corresponding conserved (anti-)BRST charges, and (ii) the infinitesimal co-BRST
and anti-co-BRST symmetry transformations and the corresponding conserved and nilpotent (anti-)co-BRST charges. 
However, we would like to lay emphasis on the fact that, as far as the physicality criteria are concerned,
it is the nilpotency property (i.e. $Q_{(a)b}^2 = Q_{(a)d}^2 = 0 $)
and the conservation laws (i.e. $\dot Q_{(a)b} = \dot Q_{(a)d} = 0 $) that are crucial for the
(anti-)BRST ($Q_{(a)b}$) and (anti-)co-BRST ($Q_{(a)d}$) charges.

There is only one missing link which we would like to highlight before we demand the physicality criteria
(i.e. $Q_{(a)b} \, |phys> = 0, \; Q_{(a)d} \, |phys> = 0$) w.r.t. the conserved and off-shell nilpotent 
(anti-)BRST and (anti-)co-BRST charges [cf. Eqs. (10),(13)], respectively. In this context, it is straightforward to
point out  that the operator forms of the infinitesimal, continuous and off-shell nilpotent 
(anti-)BRST and (anti-)co-BRST symmetry transformations
[cf. Eqs. (6),(11)] obey the following algebra
\begin{eqnarray}
\{s_b,\; s_{ab}\} = 0, \qquad \{s_b,\; s_{ad}\} = 0, \qquad \{s_d,\; s_{ad}\} = 0, \qquad \{s_d,\; s_{ab}\} = 0, 
\end{eqnarray}
which is also reflected in the algebra of the corresponding off-shell nilpotent (ant-)BRST and (anti-)co-BRST charges
[cf. Eqs. (10),(13)] because we find that the following relationships, 
amongst the above conserved charges, are {\it correct}, namely;
\begin{eqnarray}
s_b\; Q_{ab} &=& - i\, \{Q_{ab}, \; Q_b\} = 0, \qquad \quad s_d\; Q_{ad} = - i\, \{Q_{ad}, \; Q_d\} = 0, \nonumber\\
s_b\; Q_{ad} &=& - i\, \{Q_{ad}, \; Q_b\} = 0, \qquad \quad s_d\; Q_{ab} = - i\, \{Q_{ab}, \; Q_d\} = 0, \nonumber\\
s_{ab}\; Q_{b} &=& - i\, \{Q_{b}, \; Q_{ab}\} = 0, \qquad \quad s_{ad}\; Q_{d} = - i\, \{Q_{d}, \; Q_{ad}\} = 0, \nonumber\\
s_{ad}\; Q_{b} &=& - i\, \{Q_{b}, \; Q_{ad}\} = 0, \qquad \quad s_{ab}\; Q_{d} = - i\, \{Q_{d}, \; Q_{ab}\} = 0, 
\end{eqnarray}
where, in terms of the anticomutators between  the conserved and nilpotent charges, the {\it first} two lines are exactly the same as the
{\it last} two lines. The above two equations (17) and (18) are interrelated and they reflect the absolute anticommutativity 
properties of the nilpotent (anti-)BRST and (anti-)co-BRST  transformations and corresponding conserved and nilpotent charges,
respectively. Thus, ultimately, 
we define a {\it unique} bosonic symmetry transformation $s_\omega$ (with $s_\omega^2 \neq 0$) which emerges out from the underlying {\it four}
nilpotent transformations [cf. Eqs. (6),(11)]  that exist in our theory. This {\it unique} anticommutator is: 
\begin{eqnarray}
s_\omega = \{s_b,\; s_d\} = - \; \{s_{ad},\; s_{ab}\}.
\end{eqnarray}
The sanctity of the above operator relationship can be verified when it acts on the generic variable $\phi(t)$ of our theory that is described
by the Lagrangian $L_b$. The structure of (i) the above definition,  and (ii) the nilpotency and anticommutativity 
properties of the (anti-) BRST and (anti-)co-BRST transformation operators (17),
ensure that we have a beautiful algebra\footnote{This algebra is reminiscent of the algebra obeyed by the de Rham cohomological operators
of differential geometry [19,20] where the (co-)exteriors derivatives $(\delta)d$ and the Laplacian operators $\Delta$ satisfy the algebra:
$d^2 =\delta^2 = 0, \,\Delta = \{d, \, \delta\}, \, [\Delta, \, d] = [\Delta, \, \delta] =0$
which is popularly known as the Hodge algebra.}: $s_{(a)b}^2 = s_{(a)d}^2 = 0, \, s_\omega = \{s_b,\; s_d\} = - \,\{s_{ad},\; s_{ab}\}, \,
[s_\omega,\; s_{(a)b}] = 0, \; [s_\omega,\; s_{(a)d}] = 0$. It is very interesting to point out
that, under the above bosonic transformations, the (anti-)ghost variables $(\bar c)c)$ do not transform at all. Thus, one
of the characteristic features of the above bosonic transformations is the observation that the FP-ghost terms
of our (anti-)BRST and (anti-)co-BRST invariant Lagrangian $L_b$  do {\it not} transform under it.

It is worthwhile, at this juncture, to point out that the total Hilbert space of the BRST-quantized theory is an {\it inner}
product of the physical states (i.e. $|phys>$) and ghost states. This is due to the fact that, right from
the beginning, the (anti-)ghost variables $(\bar c)c$ are decoupled from the rest of the  theory because there is {\it no} interaction
between the physical variables and {\it them}. It is obvious that the operation of the (anti-)ghost variable operators
on the ghost states will be always {\it non-zero}.
It is clear now that one can perform the Hodge decomposition [19,20] for
a given physical state  in the total quantum
Hilbert space of states in terms of the conserved and nilpotent (anti-)BRST and (anti-)co-BRST charges 
and choose the {\it real} physical state (i.e. $|phys>$) to be the harmonic state [20] which will be annihilated by the  the above charges due
to the requirement of the physicality criteria (i.e. $Q_{(a)b} \, |phys> = 0, \; Q_{(a)d} \, |phys> = 0$). A close and careful look at 
the precise (and concise) expressions for the off-shell nilpotent (anti-)BRST and (anti-)co-BRST charges [cf. Eqs. (10),(13)] demonstrates that
the physicality criteria lead to the following quantization conditions, in terms of the operator forms of the first-class constraints
of our {\it classical} gauge theory, on the physical states (i.e. $|phys>$) at the {\it quantum} level, namely;
\begin{eqnarray}
 b\; |phys> = 0 \qquad &\Longrightarrow& \qquad p_\zeta \; |phys> = 0, \nonumber\\
\dot b\; |phys> = 0 \qquad &\Longrightarrow& \qquad \bigl [g\, (x\, p_y - y\, p_x) + p_z \bigr ]\; |phys> = 0,
\end{eqnarray}
which are consistent with the Dirac-quantization conditions for the theories that are endowed with any category  of constraints [12,13]. 
In the above derivation, we have used the definition of the canonical conjugate momentum 
($p_\zeta$) w.r.t. the gauge variable $\zeta (t)$
and the EL-EoM w.r.t. the above variable which leads to: $\dot b = -\;\bigl [g\, (x\, p_y - y\, p_x) + p_z \bigr ]$
that is derived from the (anti-)BRST and (anti-)co-BRST invariant Lagrangian $L_b$ [cf. Eq. (7)].

\section{Conclusions}

In our present endeavor, we have concentrated on (i) the infinitesimal, continuous and local {\it classical} gauge symmetry
transformations ($\delta_g$) that are generated by the first-class constraints of the {\it classical} FLPR model, (ii) the
{\it quantum} counterparts (of the above {\it classical} gauge transformations) as the off-shell nilpotent
and absolutely anticommuting (anti-)BRST symmetry transformations ($s_{(a)b}$) under which the total kinetic terms of
the (anti-)BRST invariant {\it quantum} Lagrangian (7) for the FLPR model remain invariant, (iii) the off-shell nilpotent and
absolutely anticommuting (anti-)co-BRST symmetry transformations ($s_{(a)d}$) under which the total gauge-fixing terms
of the (anti-)BRST and (anti-)co-BRST invariant Lagrangian $L_b$ remain
invariant, and (iv) the bosonic symmetry transformation [which is a unique anticommutator: $s_\omega = \{s_b,\; s_d\} = - \; \{s_{ad},\; s_{ab}\}$
constructed from the nilpotent (anti-)BRST and (anti-)co-BRST symmetry transformations] under which the FP-ghost terms of the Lagrangian $L_b$
remain invariant\footnote{There is a set of standard
ghost-scale continuous transformations in the theory under which {\it only} the (anti-)ghost variables $(\bar c)c$ 
transform and rest of the variables do {\it not} transform at all. However, we have {\it not} discussed this 
symmetry invariance in our present endeavor
(as we have discussed {\it all} the continuous symmetries and corresponding charges for our present theory
in our very recent  publication [8]).}. The algebraic structure of the above 
transformation operators (at the quantum level) is reminiscent of the Hodge algebra of the de Rham cohomological operators of
differential geometry [19,20]. Furthermore, we have been able to establish that the operator forms of the first-class constraints 
of the {\it classical} gauge theory annihilate the physical states (i.e. $|phys>$) at the {\it quantum} level when we demand the
validity of the physicality criteria (i.e. $Q_{(a)b} \, |phys> = 0, \; Q_{(a)d} \, |phys> = 0$) w.r.t. the
conserved and nilpotent (anti-)BRST and (anti-)co-BRST charges. We obtain the {\it same} quantization conditions 
(20) because there is {\it no true} duality in the 1D quantum mechanical system of FLPR model. In the context of the
D-dimensional ($D \geq 2$) Abelian $p$-form ($p = 1, 2, 3...$) gauge theories, we find that the physical states (i.e. $|phys>$)
are annihilated by the first-class constraints w.r.t. the (anti-) BRST charges (i.e. $Q_{(a)b} \, |phys> = 0 $) and
the dual-versions of {\it these} constraints emerge out from the requirement (i.e. $ Q_{(a)d} \, |phys> = 0 $) due
to the physicality criteria for the field-theoretic models for the Hodge theories (see, e.g. [4]).
The future perspective of the
above continuous symmetry transformations (and corresponding charges) lies in the proof that the 1D quantum 
mechanical system of the FLPR model is an example of the toy model for the Hodge theory where various discrete and continuous symmetries
(and corresponding charges) co-exist together in a meaningful manner. To be precise, we have demonstrated [8] the existence of 
a set of discrete symmetries and several continuous symmetries (and corresponding conserved charges) that
provide the physical realizations of the de Rham cohomological operators of differential geometry [19, 20] at the
algebraic level (where the set of dual discrete symmetries provide the physical realization of the Hodge duality
operation in the analogue of the mathematical relationship between the (co-)exterior derivatives).\\


\noindent
{\bf Acknowledgments}

\vskip 0.5cm

\noindent
Thanks are due to  Mr. A. K. Rao for useful discussions. A few nice comments, suggestions and thought-provoking queries
by our {\it Reviewer} are gratefully acknowledged, too.  \\

\end{document}